\documentclass{iaus} 
\usepackage{graphicx} 

\title[Upper limits on the mass of SBHs] 
{Upper limits on the mass of supermassive black holes from 
HST/STIS archival data}

\author[Corsini et al.]   
{E. M. Corsini$^1$, A. Beifiori$^1$, E. Dalla Bont\`a$^1$, 
  A. Pizzella$^1$, \break
  L. Coccato$^2$, M. Sarzi$^3$, \and F. Bertola$^1$}
 
\affiliation{$^1$Dipartimento di Astronomia, Universit\`a di Padova, 
  Padova, Italy\\[\affilskip] 
$^2$Kapteyn Astronomical Institute, University of Groningen, Groningen, 
  The Netherlands\\[\affilskip] 
$^3$Centre for Astrophysics Research, University of Hertfordshire,  
  Hatfield, UK}

\pubyear{2006} 
\volume{238}  
\pagerange{001--999} 
\date{??? and in revised form ???} 
\jname{Black Holes: from Stars to Galaxies -- across the Range of Masses} 
\editors{V. Karas \& G. Matt, eds.} 
\begin{document} 
  
\maketitle 
 
\begin{abstract} 
The growth of supermassive black holes (SBHs) appears
to be closely linked with the formation of spheroids. There is a
pressing need to acquire better statistics on SBH masses, since the
existing samples are preferentially weighted toward early-type
galaxies with very massive SBHs. With this motivation we started a
project aimed at measuring upper limits on the mass of the SBHs in the
center of all the nearby galaxies ($D<100$ Mpc) for which STIS/G750M
spectra are available in the HST archive. These upper limits will be
derived by modeling the central emission-line widths observed in the
H$\alpha$ region over an aperture of $\sim0.1''$. Here we present our
results for a subsample of 22 S0-Sb galaxies within 20
Mpc.
%
%
\end{abstract} 
 
\firstsection 
\section{Introduction} 
 
The census of supermassive black holes (SBHs) is large enough to probe
the links between mass of SBHs and the global properties of the host
galaxies (see Ferrarese \& Ford 2005). However, accurate measurements
of SBH masses ($M_\bullet$) are available for a few tens of galaxies
and the addition of new determinations is highly desirable.

To this purpose we started a project aimed at measuring upper limits
on $M_\bullet$ in the center of all the nearby galaxies ($D<100$ Mpc)
for which STIS/G750M spectra are available in the HST archive. We
retrieved data for 213 galaxies spanning over all the morphological
types. This will extend previous works by Sarzi et al. (2002) and
Verdoes Kleijn et al. (2006).
Here we analyze a subsample of 22 galaxies. They have been selected to
be S0-Sb within 20 Mpc and to have a measured stellar velocity
dispersion ($\sigma_\star$). All the galaxies were observed as part of
the HST/SUNNS project (PI: H.-W. Rix, GO-7361), except for NGC 4435
(Coccato et al. 2006).

\section{Data reduction and analysis}

The STIS/G750M spectra were obtained with the $0.2''\times52''$ slit
crossing either the galaxy nucleus along a random position angle
(SUNNS) or along the galaxy major axis (NGC 4435). The observed
spectral region includes the [N~II]$\lambda\lambda6548,6583$\AA ,
H$\alpha$, and [S~II]$\lambda\lambda6716,6730$\AA\ emission lines.
Data reduction was performed with the STIS pipeline, which we
implemented to clean cosmic rays and hot pixels. For each galaxy we
obtained the nuclear spectrum by extracting a $0.25''$-wide ($<25$ pc)
aperture centered on the continuum peak.
The ionized-gas velocity dispersion was measured by the fitting
Gaussians with the same width and velocity to the narrow component of
the [N~II] and [S~II] emission lines. The H$\alpha$ line and broad
components were fitted with additional Gaussians.

We assumed that the ionized gas resides in a thin disk and moves onto
circular orbits. The local circular velocity is dictated by the
gravitational influence of the putative SBH. To derive the upper
limits on $M_\bullet$ we first built the gaseous velocity field and
projected it onto the sky plane according to the disk
orientation. Then we observed it by simulating the actual setup of
STIS and effects related to the STIS PSF, slit width and the charge
bleeding as done by Coccato et al. (2006). Including the mass of the
stellar component would lead to tighter upper limits on $M_\bullet$.
We have no information on the orientation of the gaseous disk within
the central aperture. For this pilot project we assumed that the disk
is either nearly face on ($i=33^\circ$) or edge on ($i=81^\circ$) and
the slit is placed along its major axis. We derived the intrinsic flux
radial profile of the gaseous disk by fitting a Gaussian function to
the narrow emission-line fluxes taking into account of disk
inclination and STIS PSF.

\section{Results}

The resulting upper limits are consistent with those derived by Sarzi
et al. (2002) for the SUNNS galaxy sample and by Coccato et al. (2006)
for NGC~4435. Therefore, we are confident to obtain reliable estimates
of the upper limit on $M_\bullet$ for all the sample galaxies with
emission lines in their spectra.
This will allow to increase the statistical significance of the
relationships between $M_\bullet$ and galaxy properties and identify
peculiar objects worthy of further investigations. This is the case of
the $M_\bullet$ of NGC~3351, NGC~4314, and NGC~4143, which lie below
the $M_\bullet-\sigma_\star$ relation (Fig. 1).

\begin{figure}
\begin{center}
 \includegraphics[height=7cm]{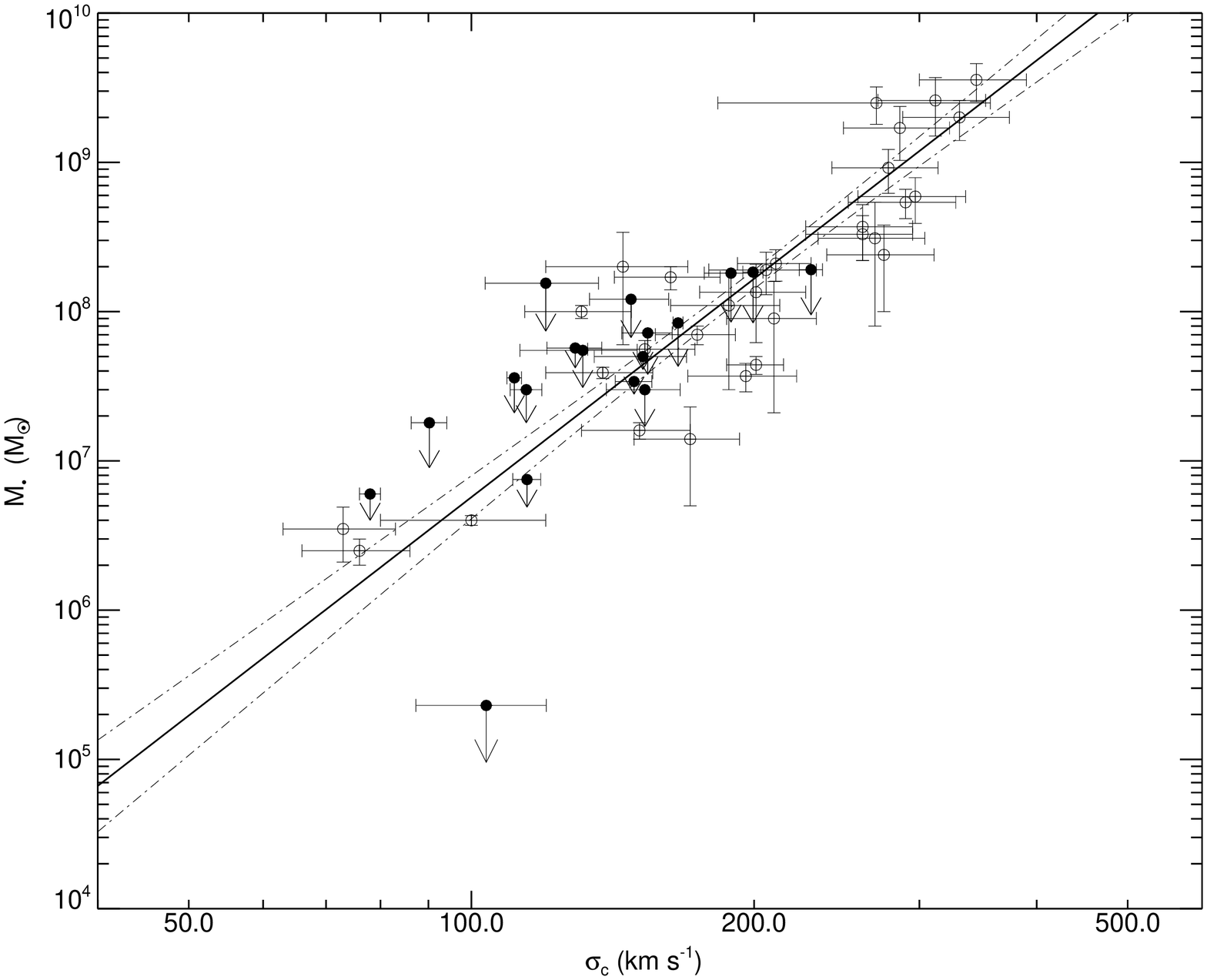}
\end{center}
  \caption{Comparison between our $M_\bullet$ upper limits (filled
   circles) and the $M_\bullet-\sigma_\star$ relation. The edges of the
   arrows correspond to upper limits obtained with $i=33^\circ$ and
   $i=81^\circ$, respectively. The open circles show the galaxies with
   accurate measurements of $M_\bullet$ (Ferrarese \& Ford 2005).}
\end{figure}

\end{document}